\documentclass[showpacs,twocolumn,preprintnumbers,amsmath,amssymb]{revtex4-1}
\usepackage[english]{babel}
\usepackage{graphicx}
\usepackage[utf8]{inputenc}

\begin{document}

\title{Atomic structure and energetics of large vacancies in graphene}

\author{J. Kotakoski}
\email{jani.kotakoski@iki.fi}
\affiliation{University of Vienna, Department of Physics, Boltzmanngasse 5,
1090 Vienna, Austria}
\affiliation{University of Helsinki, Department of Physics, P.O. Box 43,
FI-00014, Finland}
\author{F.~R. Eder}
\affiliation{University of Vienna, Department of Physics, Boltzmanngasse 5,
1090 Vienna, Austria}
\author{J.~C. Meyer}
\affiliation{University of Vienna, Department of Physics, Boltzmanngasse 5,
1090 Vienna, Austria}


\date{\today}

\begin{abstract}

We present a computational study on the topology, energetics and structural
deformations for a large number of experimentally observed defect
configurations in graphene.  We find that both the number of lost hexagonal
carbon rings and introduced non-hexagonal rings increase linearly as a function
of the vacancy order (number of missing atoms). The formation energies of the
defects increase by about 2.2~eV per missing atom after an initial offset,
establishing these defects as the lowest energy vacancy configurations studied
in graphene to date.  In addition, we find that even small point defects, which
have been until now assumed flat, cause graphene to bend out of plane when not
restricted into prohibitively confined geometries.  This effect reaches to
relative long distances even for some of the smallest defects, significantly
reducing the stress otherwise imposed on the surrounding lattice.

\end{abstract}

\maketitle

\section{Introduction} 

Two-dimensional materials, such as graphene and single layers of hexagonal
boron nitride and transition metal dichalcogenides, have provided an
unprecedent opportunity to directly image the atomic structure of various
lattice imperfections.  For example, high resolution transmission electron
microscopy (HR-TEM) studies have revealed the atomic structure of various
graphene
defects~\cite{hashimoto_direct_2004,meyer_direct_2008,warner_dislocation-driven_2012}
and grain boundary structures~\cite{huang_grains_2011,kim_grain_2011}, while
scanning tunneling microscopy (STM) has provided a complementary
view~\cite{ugeda_point_2011,ugeda_electronic_2012,tapaszto_mapping_2012}. The
energetic electrons, which are used for the imaging within TEM, have also been
utilized to intentionally create
defects~\cite{Kotakoski_Point_2011,Meyer_Accurate_2012,Robertson_Spatial_2012,Robertson_Structural_2013,Warner_Bond_2013}
and to drive their
dynamics~\cite{Kotakoski_Point_2011,kotakoski_stone-walestype_2011,Kurasch_Atom-by-atom_2012,Lehtinen_Atomic_2013}.
However, with both TEM and STM, one remains typically agnostic about any small
height variations that may occur around the defects. This is because changes in
the sub-nm range are difficult to discern based on the focus in a TEM
experiment, whereas STM measurements for supported graphene are more sensitive
to height variations of the substrate than those of graphene, and interactions
between the STM tip and a freestanding graphene can significantly alter its
shape during the measurement.~\cite{Eder_Probing_2013}

From the theoretical point-of-view, defects in graphene have offered an
intriguing playing field allowing for a multitude of different atomic
configurations via incorporation of various non-hexagonal carbon rings into the
lattice.  An overview on research of structural defects in graphene until 2011
is provided in Ref.~\cite{Banhart_Structural_2011}. Since
then, further studies have been conducted on the atomic structure of point
defects~\cite{Warner_Bond_2013,Robertson_Spatial_2012,
Robertson_Structural_2013},
dislocations~\cite{warner_dislocation-driven_2012,Lehtinen_Atomic_2013} and
grain boundaries~\cite{Rasool_Measurement_2013}.  However, few
attempts~\cite{Jeong_Stability_2008,Wang_Structural_2012} have been made to
understand the properties of realistic larger vacancy complexes in graphene.
Further, only rarely in studies of defects in graphene (with a few notable
exceptions~\cite{Ma_Stone-Wales_2009,Liu_Cones_2010,Lehtinen_Atomic_2013}),
have out-of-plane variations been adequately addressed.  Instead, the
structures have been typically assumed flat, and the atomic projections in the
TEM images have been interpreted as corresponding to a flat structure under
significant negative strain~\cite{Warner_Bond_2013,Rasool_Measurement_2013}.
However, by bending out from the flat configuration, a thin membrane can
locally reduce the stress around defects.

In this work, we study a set of defect configurations in graphene, produced
and observed during a TEM experiment under a 100~kV electron beam.  The
formation of these defects has been previously described in
Refs.~\cite{Kotakoski_Point_2011,Meyer_Accurate_2012}. TEM images of the
defects can be found in the supplementary video S4 of
Ref.~\cite{Meyer_Accurate_2012}. We find that, on average, 1.6 hexagons will be
transformed into 1.0 non-hexagonal carbon rings per one missing atom. All of
the studied structures are observed to deform graphene in the out-of-plane
direction, which significantly reduces the local stresses around the defects.
The formation energies increase linearly as a function of the vacancy-order (by
$\sim 2.19$~eV per missing atom after an initial offset), establishing this set
of defects as having the lowest formation energies for any vacancy complexes
hitherto studied in graphene.

\section{Methods and results}

\begin{figure*}[!]
\includegraphics[width=0.8\linewidth]{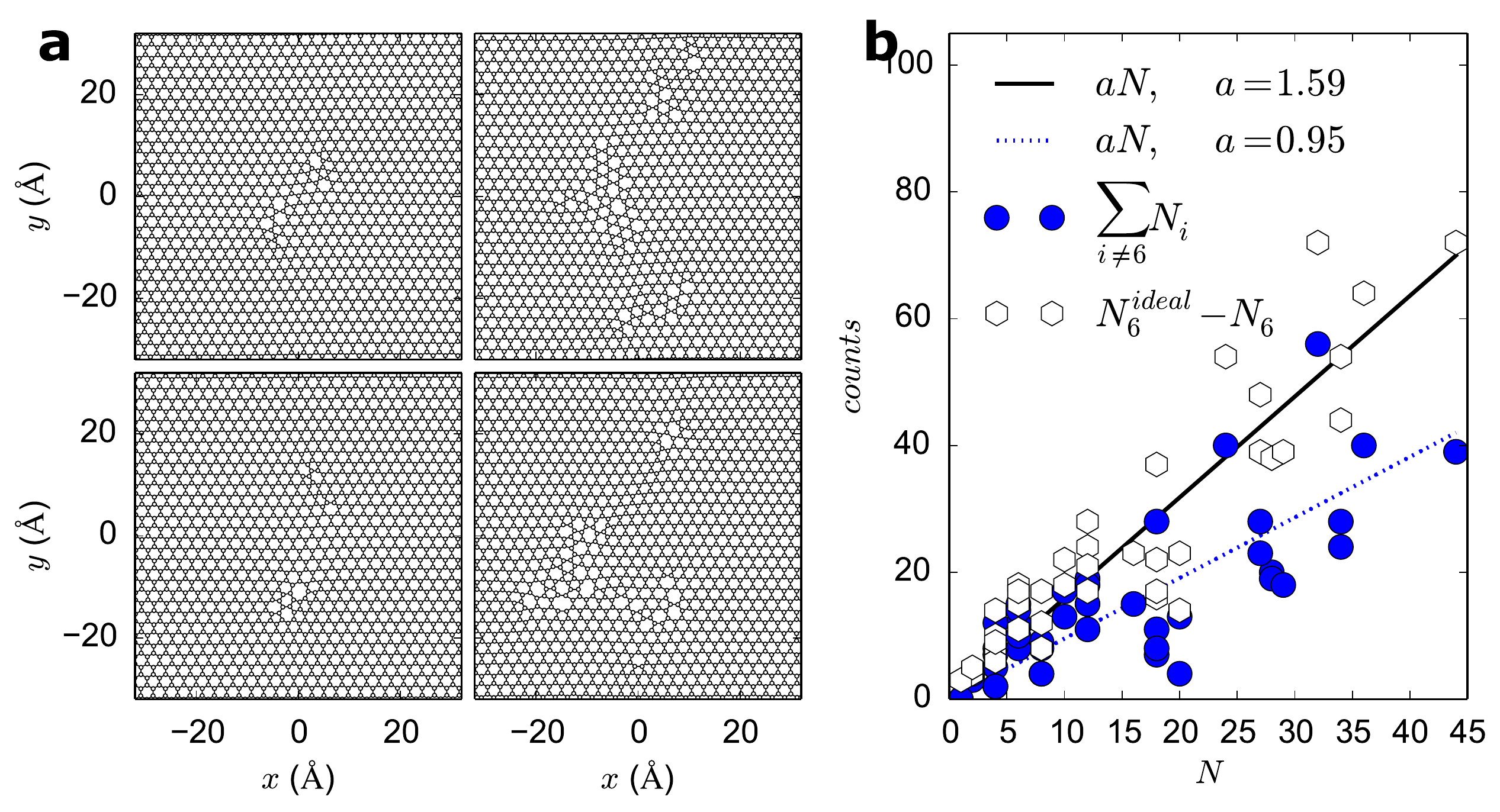}
\caption{(Color online) (a) Four examples of the 44 vacancy structures used in
this study with varying number of missing atoms ($N$). $x$ and $y$ are the
cartesian coordinates in the in-plane direction. (b) Number of missing
hexagonal carbon rings ($N_6^{ideal}-N_6$) and number of all other polygons
($\sum_{i\neq 6}N_i$) as a function of $N$. Note that for many $N$, there are
several different configurations. Lines are fits to the data.}
\label{fig::rings}
\end{figure*}

We start the analysis by looking at the topology of the defects (see
Fig.~\ref{fig::rings}a for four example structures). In Fig.~\ref{fig::rings}b,
we plot the number of introduced non-hexagonal polygons and the number of
removed hexagons (with respect to the pristine lattice) as a function of the
vacancy order (number of missing atoms). Out of all non-hexagonal carbon rings,
0.8\% were tetragons, 55.3\% pentagons, 38.2\% heptagons and 5.8\% octagons.
The number of removed hexagons increases (on average) linearly with the number
of removed atoms with a rate of ca.  1.6 per atom. This is comparable to value
of 2 for the V$_2$(5-8-5), but somewhat lower than those for the V$_1$(5-9) or the
energetically more stable divacancies~\cite{Banhart_Structural_2011}, namely,
the V$_2$(555-777) and the V$_2$(5555-6-7777), which have values of 3, 3.5 and
4.5 per missing atom, respectively. The number of non-hexagonal rings, on the
other hand, increases at a rate of about 1.0 polygons/missing atom. Comparing
to the average value of $\sim 0.6$, for all of the above-mentioned divacancies,
the difference between removed hexagons and introduced other polygons is
exactly 0.5 per missing atom (1.0 for the single vacancy). We stress that all
of the above analysis is based on the average properties of all of the defect
structures, which means that for any specific defect, the values can deviate
from the ones listed. All of the analyzed atomic structures are available
through Ref.~\cite{suppl}.

The largest defect structure considered in this study has 44 missing
atoms and involves nearly 80 lost hexagons with up to 40 other polygons
introduced. Thus, in order to contain each of the structures within a
pristine lattice of the same size, we had to create a relatively large
simulation system into which the defects were then (separately) introduced. As
a consequence, also the small well-studied defects were in this study relaxed
in an atypically large system.  We carried out structural optimization for all
of the defect structures using the conjugate gradient energy minimization
scheme employing two different analytical interaction models to describe the
energetics and inter-atomic forces in the system, namely the PII
parametrization by Brenner from Ref.~\cite{Brenner_Empirical_1990} and the
improved version of the potential (AIREBO) by Stuart et
al.~\cite{stuart_reactive_2000}. Both potentials reproduced all of the trends
reported within this work. However, since the bond lengths and formation
energies predicted by AIREBO are closer to density functional theory (DFT)
values, as will be to some extent described below, all of the results shown
were obtained with AIREBO. These simulations were carried out with the LAMMPS
code~\cite{Plimpton_Fast_1995,_lammps_????}.  The DFT results were calculated
with VASP~\cite{kresse_efficiency_1996,kresse_efficient_1996} using projector
augmented wave potentials~\cite{blochl_projector_1994}.  Plane wave cut-off was
set to 300~eV, and the exchange and correlation were described with the
generalized gradient approximation parametrized by Perdew, Burke and
Ernzerhof~\cite{perdew_generalized_1996}. Only one $\mathbf{k}$-point was used
($\Gamma$).  These parameters were dictated by the relative large system size
(up to 880 atoms for the DFT simulations).

\begin{figure}[!]
\includegraphics[width=\linewidth]{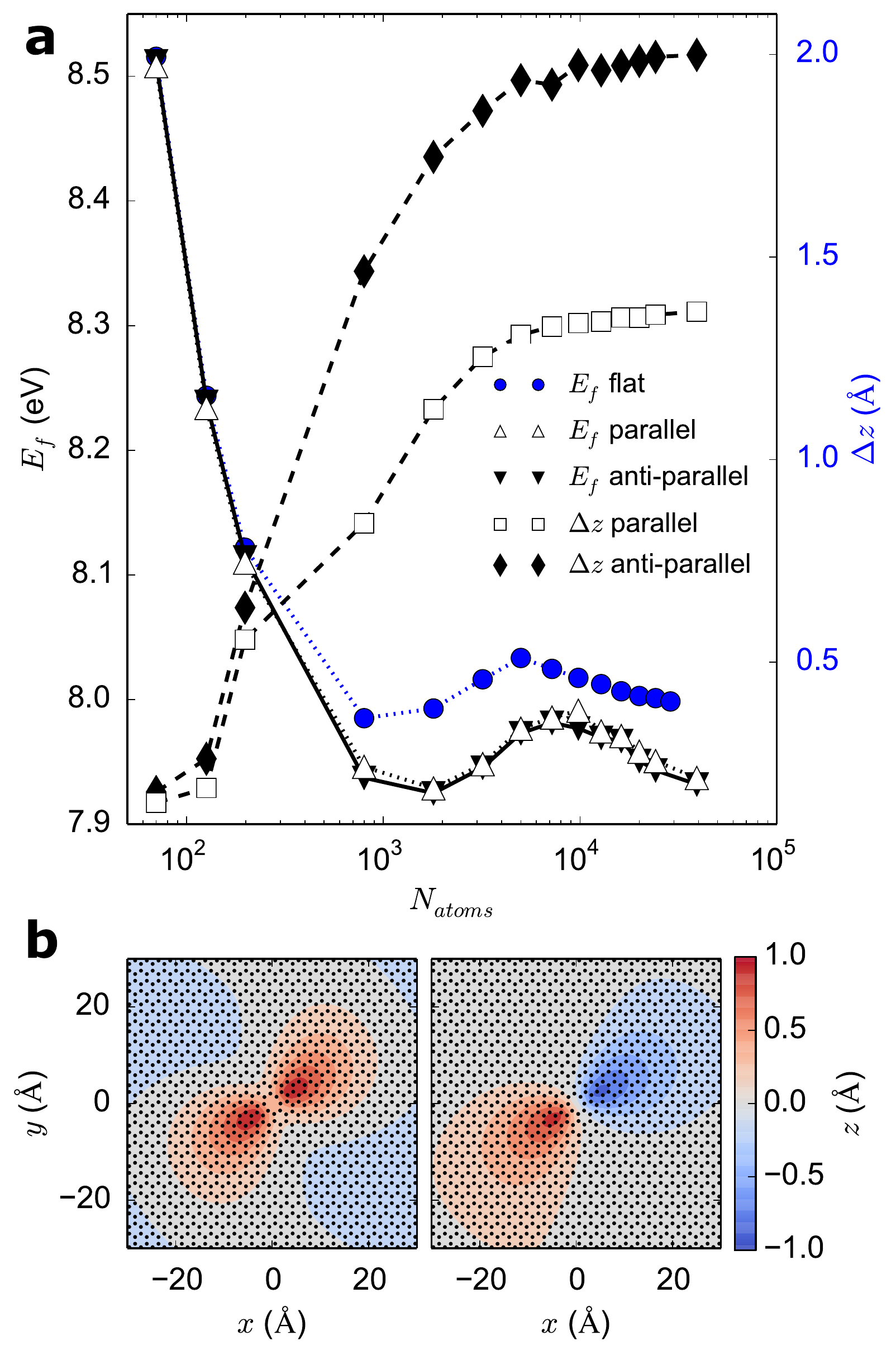}
    \caption{(Color online) Effect of system size and out-of-plane corrugations
on the formation energy of a V$_2$(5-8-5) divacancy structure in graphene. (a)
Formation energy $E_f$ as a function of the system size ($N_{atoms}$, number of
atoms) for the flat and the two non-flat configurations (with parallel and
anti-parallel symmetries) and the maximum distance of two atoms within the
relaxed non-flat structures in the out-of-plane direction ($\Delta z$). (b)
Out-of-plane corrugations for the two non-flat configurations. The lines are
guides to the eye. $x$ and $y$ are the cartesian coordinates in the in-plane
direction.} \label{fig::585} \end{figure}

During initial relaxation of the structures with both analytical potentials, we
noticed that the defects caused the lattice to bend out from the flat
configuration. In order to check that this observation was not caused by a
simulation artefact, we selected the V$_2$(5-8-5) divacancy to study this
effect. This defect has been extensively studied in the past (see, f.ex.,
Ref.~\cite{krasheninnikov_bending_2006}). Curiously, it is known to lead to a
local curvature change in carbon
nanotubes~\cite{krasheninnikov_bending_2006,kotakoski_energetics_2006}, but for
graphene, as far as we know, only flat configurations have hitherto been
reported. As can be seen in Fig.~\ref{fig::585}a, the formation energy of this
defect decreases with increasing system size up to about 1,000 atoms due to
repulsion between the defect and its mirror images over the periodic boundaries
at shorter inter-defect distances. At system sizes exceeding 100 atoms, the
flat configuration becomes less favorable than the buckled one, and remains so
up to the largest studied systems.  From this data, we can conlude that
defect-to-defect interactions reach up to at least 15~nm (corresponding to the
system size for 20,000 atoms in this study) even in the case of small point
defects in graphene.  Interestingly, we find two possible buckled structures
for this defect with an anti-parallel and a parallel symmetry with respect to
the graphene plane [see Fig.~\ref{fig::585}b]. Both of these configurations
are favored over the flat structure.

We point out that no restriction was required to keep the structure flat during
the relaxation at any system size. According to the analytical potentials, this
configuration presents thus a metastable defect state even when it is less
favored than the nonflat structure. However, one can reasonably assume that in
any practically relevant situation, the graphene membrane will tend towards the
energetically favored configuration due to external influences (f.ex., thermal
vibrations).  We assume that these reasons have until now hidden the corrugated
nature of graphene with point defects in theoretical studies.  In contrast to
rather large system sizes required to reveal the out-of-plane bending of the
divacancy, earlier quantum monte carlo and DFT results have shown that the
Stone-Wales defect bends the graphene sheet already at much smaller system
sizes (down to just a few tens of atoms)~\cite{Ma_Stone-Wales_2009}.

Also in Fig.~\ref{fig::585}a (second $y$-axis), we plot the maximum distance
for any two atoms in the out-of-plane direction for the optimized locally
buckled configurations ($\Delta z$).  The local height of the membrane around
the defect increases continuously indicating that the further decrease in
strain, mediated by the out-of-plane deformations, does not significantly alter
the energetics of the defect and that the defects become sufficiently spatially
isolated at the largest system sizes.

We further looked into how the out-of-plane relaxation correlates with the
release of the local negative strain. The length of the shortest bond in each
relaxed structure was found to increase monotonously with the defect height for
the buckled structures (from 2.4\% shrinkage with respect to ideal graphene up
to 1.7\% shrinkage over the studied range), whereas for the flat structure the
shortest bond remained at 2.4\% shrinkage regardless of the system size. The
number of bonds with more than 0.5\% shrinkage saturated at 26 for the non-flat
structures and at 36 for the flat ones at the largest system sizes. These
numbers clearly show that the out-of-plane relaxation is an important way for
the graphene lattice to relieve negative strain caused by the formation of
defects. We also checked how the deformation reacts to strain in the lattice.
It turns out that 0.5\% is enough to make the divacancy structure flat for a
20,000-atom system. However, strains up to several \% are required to flatten
any of the larger defects. A more in-depth study of these deformations is out
of the scope of the present work.

Based on the above analysis, we established 20,000 atoms as a reasonable system
size for the AIREBO simulations (we tested system sizes up to 180,000 to check
that there are no significant changes at formation energies even for the
largest defects). However, structures this large are clearly beyond what can be
typically modeled with any first principles method, $\sim 1,000$ atoms being
more typical. A comparison of our AIREBO results for systems of 20,000 atoms
and 880 atoms showed that the formation energies of small defects ($\sim 6$
missing atoms) were already accurate within 0.5~eV for the small system with a
few exceptions being accurate within 0.1~eV, while larger defects exhibited
deviations in the range of $2-3$~eV. Additionally, as was shown above, the
buckled structures appear energetically favored over the flat ones already for
the 880-atom system. (Although, the preferred mode of bending for an isolated
defect may remain hidden.)

In order to qualify our AIREBO simulations, we repeated some of the
V$_2$(5-8-5) simulations with DFT for system sizes of 72, 200 and 880 atoms.
For the 880-atomic system, we obtained formation energy of 7.13~eV for the
buckled structure (anti-parallel) and 7.21~eV for the flat one, confirming our
AIREBO results with a similar energy difference (0.04~eV for AIREBO for the
same system size).  The value for the flat structure is in line with the
results reported in the literature~\cite{Banhart_Structural_2011}.  (The AIREBO
overestimates $E_f$ for this defect by about 1~eV.) The maximum atom-to-atom
distance in the out-of-plane direction ($\Delta z$), as obtained from DFT
simulations is ca.  0.76~{\AA}, indicating that AIREBO slightly overestimates
the magnitude of the deformations. For both of the smaller systems (72 and 200
atoms), DFT simulations converged into the flat structure, showing that they
remain too small to accompany the out-of-plane relaxation.

\begin{figure}[!]
\includegraphics[width=\linewidth]{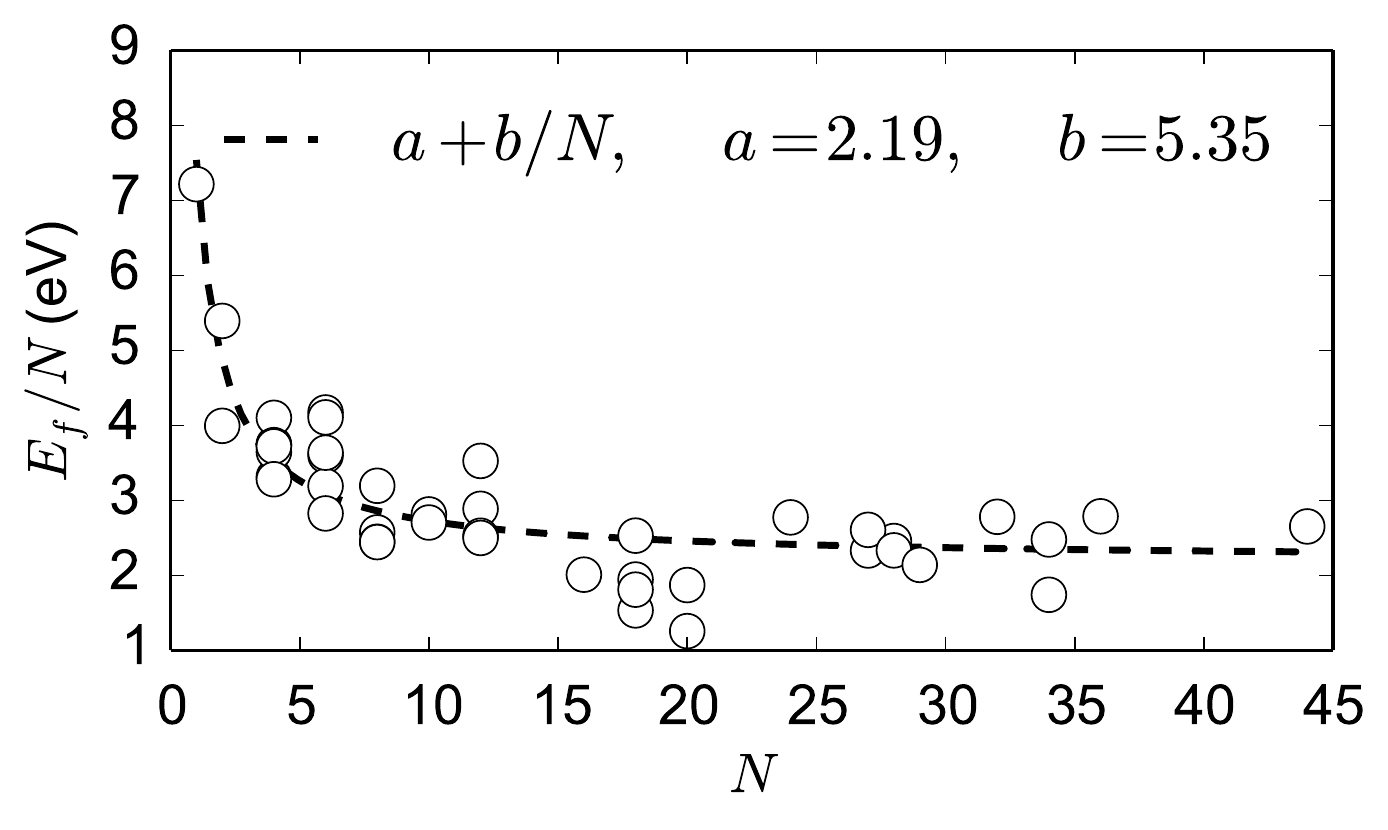}
		\caption{Formation energies per number of missing atoms $E_f/N$ for vacancy
structures, as calculated with the AIREBO potential for a system of 20,000
atoms. The dashed curve is a fit to the data.} \label{fig::Ef}
\end{figure}

After establishing the reliability of the AIREBO results regarding
energetics and structural deformations, we expanded the formation energy study
for all of the defect structures obtained from TEM images. The results are
shown in Fig.~\ref{fig::Ef} (notice that there are several different defects
with same number of missing atoms, $N$). The formation energies
scale linearly with the increasing vacancy order ($E_f/N \propto 1/N$), increasing
by ca. 2.19~eV per missing atom, with an initial offset of about 5.35~eV. This
energy penalty is associated with the local deformation of bonds 
around the locations of removed atoms.  The linear dependency is similar to
that found by Jeong et al.~\cite{Jeong_Stability_2008} for dislocation lines
and local h{\ae}ckelite structures [composed of merged V$_2$(555-777) defects]
up to 12 missing atoms, although the formation energies of the vacancy structures
considered here are substantially lower (by about 0.5~eV per missing atom).
Introduction of the first defect into the pristine lattice is associated with
the highest energy penalty (the initial offset in $E_f$), because the growing
defect can easier accumulate the non-ideal bonds due to effects like
overlapping strain fields with opposing signs~\cite{kotakoski_energetics_2006},
which leads to lower energy penalty for the removal of the
additional atoms. This behavior is the physical reason why graphene can be
turned into an amorphous 2D carbon glass via introduction of a growing number
of defects~\cite{Kotakoski_Point_2011,eder_journey_2014}.

\begin{figure*}[!]
\includegraphics[width=.9\linewidth]{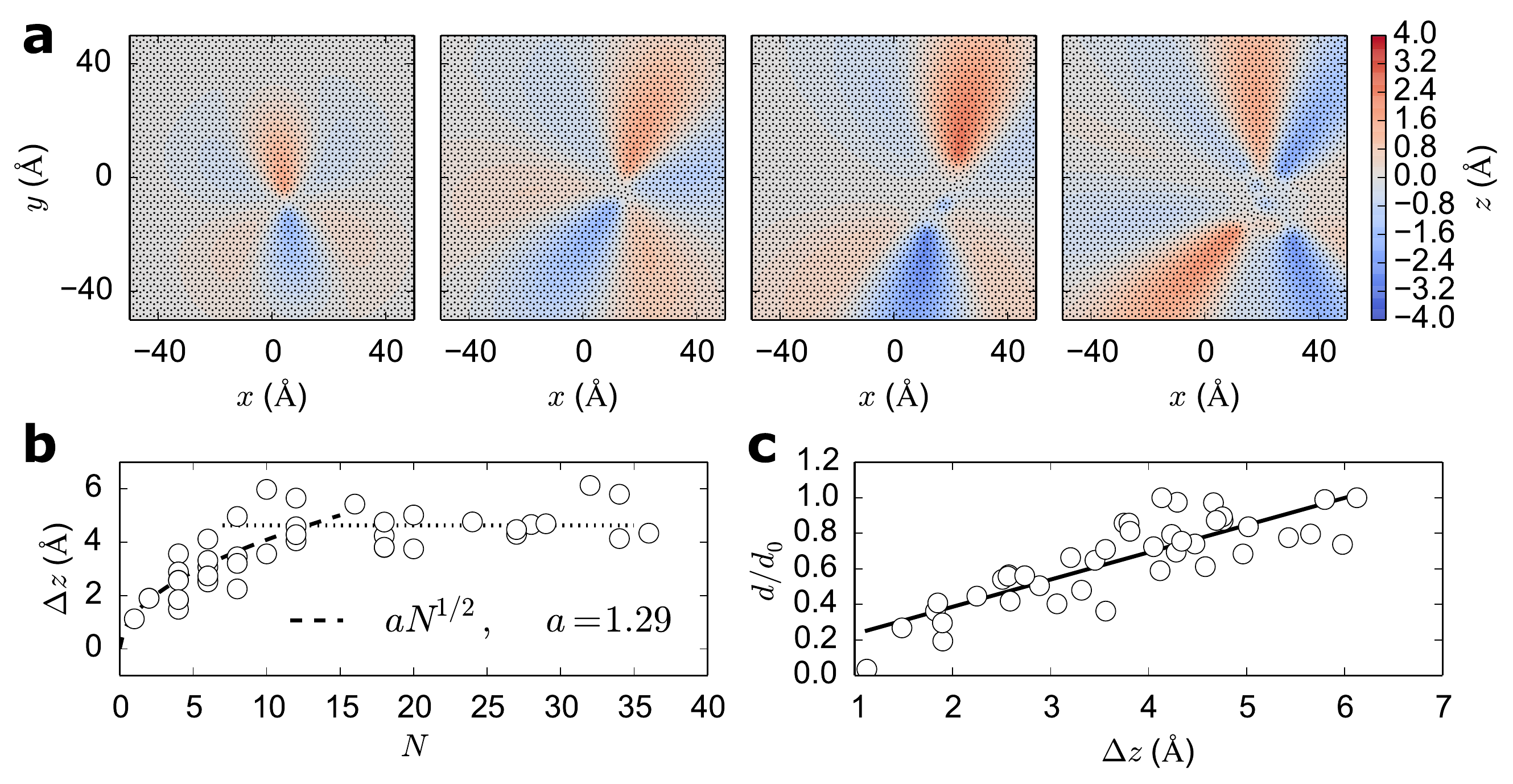}
		\caption{(Color online) Out-of-plane buckling.
(a) Height maps for four example defects with different number of
missing atoms ($N = 4, 8, 16$ and $28$, from left to right). $x$ and $y$ are
the cartesian coordinates in the in-plane direction. (b) Maximum difference in
the out-of-plane direction ($\Delta z$) for each of the defects
as a function of the number of missing atoms, $N$. The dashed and dotted lines
are fits to the data for $N<15$ and $N>15$, respectively. (c) Relative increase
of the in-plane size of the local out-of-plane deformation as compared to the
largest one as a function of $\Delta z$.}
\label{fig::defl}
\end{figure*}

Finally, we turn to look at the out-of-plane deformation caused by the larger
vacancies. As an example, four height maps for defects with different $N$ are
presented in Fig.~\ref{fig::defl}a. Also the $\Delta z$ and an estimation of the
in-plane size of the out-of-plane deformation for all defects are plotted in
Fig.~\ref{fig::defl}b,c. This estimation was made by plotting height maps, similar
to those in Fig.~\ref{fig::defl}a but for a larger area, and by drawing a circle
around each defect so that all areas with a height deviation more than
an arbitrarily selected $\Delta z_0$ were contained inside the circles. The
diameter of these circles was then used as the measure for the size. This approach
resulted in estimation of deformation length in the in-plane direction, which
is linearly dependent on $\Delta z$, showing that each of the buckled
structures has approximately the same local curvature. $\Delta z$ itself appears
to increase for defects with $N<15$ proportional to $\sqrt{N}$, which
agrees with the assumption that the area of the defect grows 
linearly with $N$. At $N\approx 15$, however, $\Delta z$ saturates to approximately a
constant value of ca. 4.3~{\AA}. This is most likely due to the fact that at
this point the growth of the defect leads to buckling at the defect itself
(increasing the frequency of the out-of-plane deformations around the defect)
without further contribution to the buckling amplitude of the membrane. This
can be seen as the increased number of ripples for the largest defect in Fig.~\ref{fig::defl}a
as compared to the smaller defects. These trends, and even absolute values,
remained almost completely unchanged also for the largest studied system sizes
(up to 180,000 atoms). The actual interaction distances reach up to tens of nm,
as was already discussed above in the case of V$_2$(5-8-5).

\section{Conclusions} 

As a conclusion, we have presented a computational study on the topology,
energetics and atomic structure of experimentally observed defect structures in
graphene. Unlike what has been previously assumed, all of the defects, including the
smallest ones, cause local buckling of graphene to lower the negative strain
otherwise imposed on the lattice. The number of removed hexagonal carbon rings
and introduced other polygons are shown to systematically increase as a
function of the number of missing atoms. The ratios of the other polygons
remain approximately constant. Formation energies of the defects increase
linearly with the increasing defect size by ca. 2.2~eV per removed atom after
an initial onset of about 5.4~eV. Finally, the height of the buckling is shown
to increase as a square root of the number of removed atoms until 15 missing
atoms before saturating to up to 4~{\AA} for the largest defects. These
deformations result in an interaction length between defects in graphene in the
regime of tens of nanometers.

We acknowledge Austrian Science Fund (FWF): M~1481-N20 and Helsinki University
Funds for funding and the Vienna Scientific Cluster for generous
grants of computational resources.



\end{document}